\documentclass[aps,prb,twocolumn,superscriptaddress,a4paper,longbibliography]{revtex4-2}
\usepackage[dvipdfmx]{graphicx}
\usepackage{color}
\usepackage{amsmath}
\usepackage{amssymb}
\usepackage{lipsum}
\begin{document}

\title{Quantum skyrmion dynamics studied by neural network quantum states}

\author{Ashish Joshi}
\affiliation{Department of Physics, Kyoto University, Kyoto 606-8502, Japan}
\email[]{joshi.ashish.42a@st.kyoto-u.ac.jp}
\author{Robert Peters}
\affiliation{Department of Physics, Kyoto University, Kyoto 606-8502, Japan}
\author{Thore Posske}
\affiliation{I. Institute for Theoretical Physics, Universit\"at Hamburg, Notkestraße 9, 22607 Hamburg, Germany}
\affiliation{The Hamburg Centre for Ultrafast Imaging, Luruper Chaussee 149, 22761 Hamburg, Germany}
\begin{abstract}

We study the dynamics of quantum skyrmions under a magnetic field gradient using neural network quantum states. First, we obtain a quantum skyrmion lattice ground state using variational Monte Carlo with a restricted Boltzmann machine as the variational ansatz for a quantum Heisenberg model with Dzyaloshinskii-Moriya interaction. Then, using the time-dependent variational principle, we study the real-time evolution of quantum skyrmions after a Hamiltonian quench with an inhomogeneous external magnetic field. We show that field gradients are an effective way of manipulating and moving quantum skyrmions. Furthermore, we demonstrate that quantum skyrmions can decay when interacting with each other. This work shows that neural network quantum states offer a promising way of studying the real-time evolution of quantum magnetic systems that are outside the realm of exact diagonalization.

\end{abstract}

\maketitle

\section{Introduction}
\label{introduction}

Manipulating magnetic structures and understanding their dynamics is crucial for their potential use in spintronics devices. Among these magnetic structures, magnetic skyrmions have received great attention in recent years because of their topological protection and ease of motion, making them an attractive candidate for magnetic storage devices like skyrmion-based racetrack memory \cite{nagaosa_topological_2013,bogdanov_physical_2020,kiselev_chiral_2011,romming_writing_2013,schutte_inertia_2014,tomasello_strategy_2014,heinze_spontaneous_2011}. These quasiparticles can be manipulated by various methods like spin and charge currents \cite{jiang_direct_2017,juge_current-driven_2019,woo_current-driven_2018,yu_skyrmion_2012,jonietz_spin_2010}, electric and magnetic field gradients \cite{upadhyaya_electric-field_2015,liu_skyrmion_2013,zhang_manipulation_2018,komineas_skyrmion_2015}, temperature gradients \cite{everschor_rotating_2012,kong_dynamics_2013} and microwaves \cite{wang_driving_2015}. The size of magnetic skyrmions can range from micrometers to a few times the atomic lattice spacing \cite{nagaosa_topological_2013}. 
However, the properties of magnetic skyrmions are mainly analyzed classically, which may only be relevant for large skyrmions.
%\textcolor{red}{(Let's stress: Not only the dynamics but also other stability and other properties are mainly analyzed classical.}
For example, while classically, the dynamics of skyrmions are described by the Landau-Lifshitz-Gilbert equation, small-sized skyrmions cannot be described using classical spins as quantum effects can play an important role.

Only recently, the static properties of quantum skyrmions have been studied in systems with Dzyaloshinskii-Moriya interaction (DMI) \cite{takashima_quantum_2016,gauyacq_model_2019,siegl_controlled_2022,haller_quantum_2022,sotnikov_probing_2021,maeland_quantum_2022,joshi_ground_2023}, frustration \cite{lohani_quantum_2019}, itinerant magnetism \cite{Kobayashi2022} and with $f$-electron systems \cite{peters_quantum_2023}. The presence of DMI or frustration makes quantum skyrmions challenging to study numerically using quantum Monte Carlo methods. While most works have focused on small systems amenable to exact diagonalization, a few works have tackled larger lattices using the density matrix renormalization group \cite{haller_quantum_2022} and neural network quantum states \cite{joshi_ground_2023}. However, research in the dynamical properties of quantum skyrmions is lacking as it is considerably more challenging numerically. A very recent work showed the onset of a quantum skyrmion Hall effect in $f$-electron systems under linear response theory using dynamical mean field theory \cite{peters_quantum_2023}. To shed more light on the dynamics of quantum skyrmions, a full nonequilibrium calculation on large lattices is needed.

Variational methods based on artificial neural networks offer a feasible approach to approximate the static and dynamic properties of quantum many-body systems \cite{carleo_solving_2017,nomura_restricted-boltzmann-machine_2017,choo_symmetries_2018,saito_machine_2018,szabo_neural_2020,hibat-allah_recurrent_2020,kochkov_learning_2021,inui_determinant-free_2021,viteritti_transformer_2023,donatella_dynamics_2023,sinibaldi_unbiasing_2023}. These methods use an artificial neural network to represent the variational wave function, known as a neural network quantum state (NQS), which learns the target state using a gradient-based optimization scheme. NQS-based methods are gaining popularity because of their higher expressive capacity than conventional methods and their ability to simulate large lattices in high dimensions \cite{sharir_neural_2022,sharir_towards_2022,moreno_fermionic_2022}. State-of-the-art results have been obtained by combining NQS with variational Monte Carlo (VMC) for ground state calculations \cite{roth_high-accuracy_2023,rende_simple_2023} and with time-dependent variational Monte Carlo (t-VMC) for real time evolution \cite{schmitt_quantum_2020,mendes-santos_highly_2023}. In the context of quantum skyrmions, we showed in our previous work that an NQS can efficiently represent the quantum skyrmion ground state at medium and strong DMI \cite{joshi_ground_2023}.

In this work, we study the real-time evolution of a quantum skyrmion lattice in the presence of an external magnetic field gradient using NQS and t-VMC. First, we obtain a quantum skyrmion lattice as the ground state of a two-dimensional spin-$1/2$ Heisenberg Hamiltonian with DMI. The spins in this quantum skyrmion lattice have nonzero quantum entanglement. Then, we quench the Hamiltonian with a nonuniform external magnetic field and evolve the system according to the time-dependent Schr{\"o}dinger equation using t-VMC. We show that quantum skyrmions move diagonally to the field gradient, resembling a skyrmion Hall effect, with a velocity that is larger in the direction perpendicular to the magnetic field gradient. 
%\textcolor{red}{(Maybe we should delete Hall-like. There is no applied current. We do not know how current are induced after the quench. If we want to mention the Hall effect, how about (resembling the Hall effect) "moving diagonally to the field gradient resembling a Hall effect"? )}
The quantum skyrmions interact with each other, leading to the formation of an exceptional configuration with the topological charge of a meron, which causes the decay of a quantum skyrmion. Merons and antimerons are vortex-like spin textures that are quantized to half the skyrmion number $N$, a topological invariant used to characterize skyrmions (Eq.~\eqref{sk_no}).
%\textcolor{red}{To our knowledge, we report the first observation of a quantum }\textcolor{red}{antimeron formed due to the interaction of quantum skyrmions. }. Merons and antimerons have been observed to emerge from the interactions of skyrmions previously in the presence of magnetic fields \cite{yu_transformation_2018} and microwaves \cite{eto_dynamical_2021}. 
%\textcolor{red}{(Here we need a little bit more about our Merons. More than just saying we have quantum merons!)} 
Our work shows that NQS can be used as a variational ansatz to study the ground state and nonequilibrium properties of quantum skyrmions with system sizes that are not feasible using exact methods.

The paper is organized as follows: In Sec.~\ref{model}, we describe the model used in our simulations. In Sec.~\ref{method}, we discuss the NQS-based variational methods to obtain ground states and to perform real-time evolution. The details of various parameters in our calculations are described in Appendix~\ref{appendix_A}. In Sec.~\ref{ground_state}, we discuss the properties of the quantum skyrmion ground state. In Sec.~\ref{dynamics_of_quantum_skyrmions}, we study the dynamical properties of quantum skyrmions. Finally, we summarize our work in Sec.~\ref{summary}.

\section{Model}
\label{model}
\begin{figure}
    \centering
    \includegraphics[width=\linewidth]{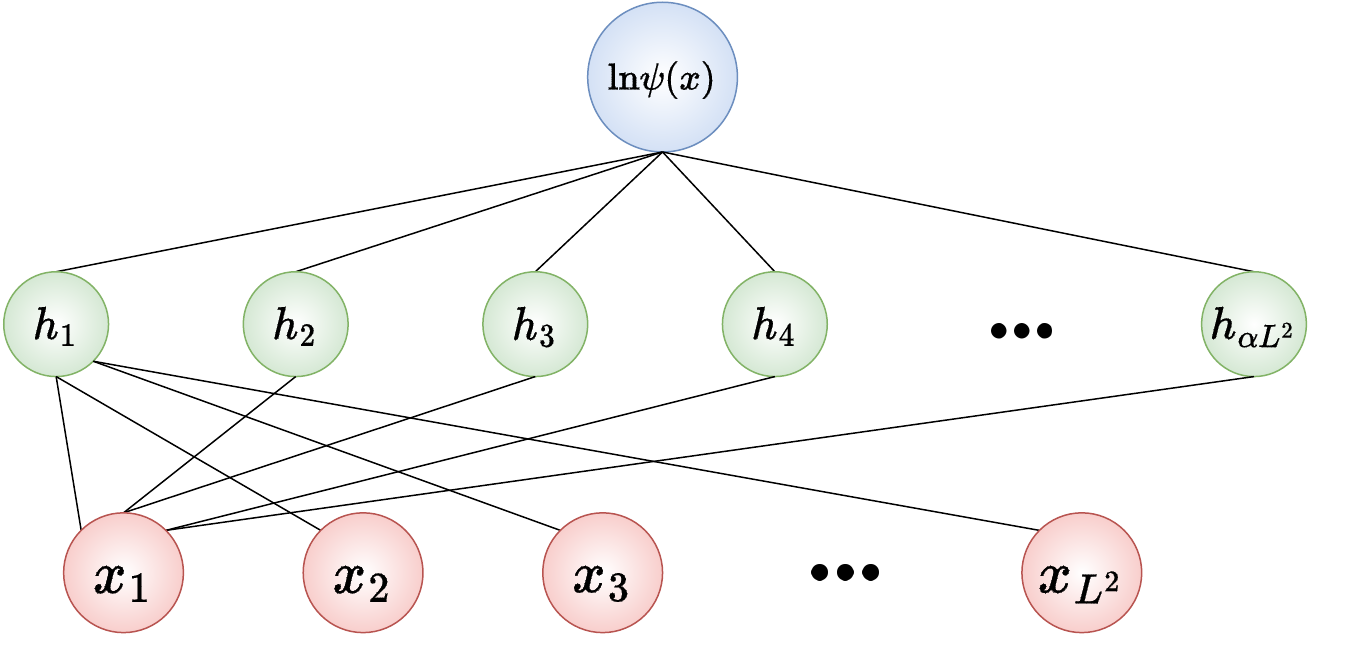}
	\caption{Restricted Boltzmann machine used as the neural network quantum state. The inputs are the spin configurations in $\sigma^z$ basis, and the output is the logarithm of the wave function (Eq.~\eqref{NN_eqn}). The hidden layer contains $\alpha L^2$ neurons, where $\alpha=2$ in our case.}
	\label{RBM}
\end{figure}

We study the spin-$1/2$ Heisenberg Hamiltonian with DMI and anisotropy on a two-dimensional lattice with periodic boundary conditions. The Hamiltonian is given as 
\begin{equation}
    \begin{split}
        H_0=&-J\sum_{\left<ij \right>}(\sigma_i^x\sigma_j^x+\sigma_i^y\sigma_j^y) - A\sum_{\left<ij \right>}\sigma_i^z\sigma_j^z\\ 
        &-D\sum_{\left<ij \right>}(\boldsymbol{u}_{ij}\times\hat{\boldsymbol{z}})\cdot({\boldsymbol{\sigma}}_i\times\boldsymbol{{\sigma}}_j)+B^z\sum_i\sigma_i^z.
    \end{split}
    \label{Ham}
\end{equation}
Here, $J$ is the Heisenberg exchange term, $A$ is the Heisenberg anisotropy term, $D$ is the strength of the DMI, and $B^z$ is the strength of the homogeneous external magnetic field. We take $\hbar=1$. The Pauli matrices are denoted by $\boldsymbol{\sigma}=\{\sigma^x,\sigma^y,\sigma^z\}$ and $\boldsymbol{u}_{ij}$ is the unit vector from site $i$ to site $j$. The first three terms are summed over the nearest neighbors denoted by $\langle ij \rangle$. A quantum skyrmion state can emerge due to the competition between the ferromagnetic exchange term and the noncolinear DMI term, stabilized by the anisotropy and the external magnetic field.

\begin{figure*}
	\centering
	\includegraphics[width=\textwidth]{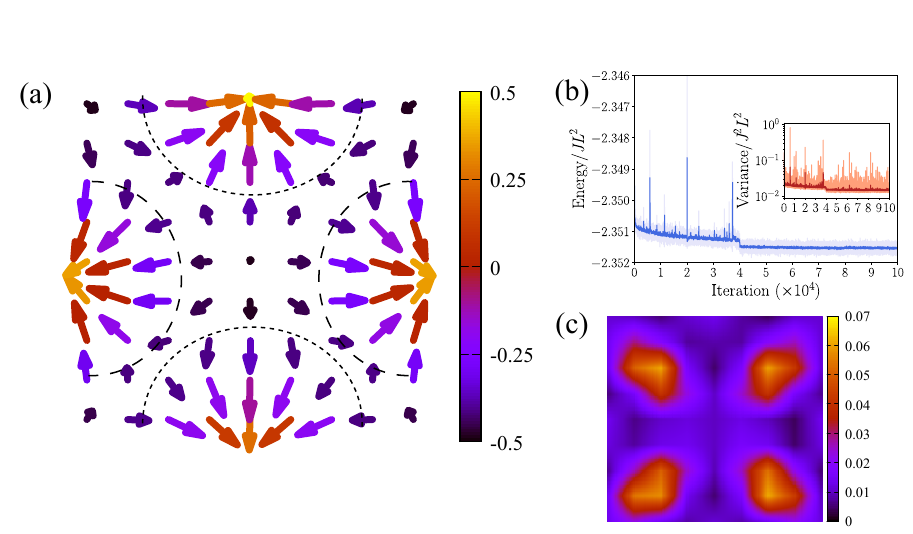}
	\caption{Quantum skyrmion lattice ground state of Eq.~\eqref{Ham} at parameters $D=J$, $A=0.5J$ and $B^z=J$. (a) Ground state spin expectation values, showing a skyrmion lattice with two quantum skyrmions. The color map indicates the $z$ component of the spin expectation value. (b) Convergence of variational energy per spin over the number of iterations for a $9\times 9$ lattice, with the energy variance per spin in the inset. The lighter color shows the values at each iteration, while the darker color shows the moving average over $30$ iterations. (c) Renyi entropy of the ground state. The entropy is largest in the space between two skyrmions.}
	\label{gs_fig}
\end{figure*}

\section{Method}
\label{method}

To obtain the ground state of the above Hamiltonian, we use variational Monte Carlo with neural network quantum states (NQS) as the variational ansatz. The many-body wave function is approximated using an artificial neural network that encodes the complex-valued coefficients $\psi_{\boldsymbol{\theta}}(x)$,
\begin{equation}
    \left|\psi_{\boldsymbol{\theta}}\right>=\sum_{x}\psi_{\boldsymbol{\theta}}(x)\left|x\right>.
\end{equation}
Here, $\boldsymbol{\theta}$ are the variational parameters, and $\left|x\right>$ are the local basis states, which in our case are the eigenvalues of the $\sigma^z_j$ operators. We use a restricted Boltzmann machine (RBM) with complex weights and biases as the variational wave function. The RBM consists of an input layer that takes the spin configurations $\left|x\right>$ as input and a hidden layer with variational parameters $\boldsymbol{\theta}=(a,W,b)$, see Fig.~\ref{RBM}. Here, $a$ are input biases, and $W$ and $b$ are hidden weights and biases, respectively. The length of one side of the lattice is given by $L$, and $\alpha$ is the hidden unit density. In this study, $\alpha=2$ is used for both ground state and time evolution calculations. Increasing $\alpha$ increases the expressiveness of the network, resulting in better accuracy but with higher computational cost. The output is the logarithm of the unnormalized wave function
\begin{equation}
    \text{ln}(\psi_{\boldsymbol{\theta}}(x))= \sum_i^{\alpha L^2}a_ix_i + \text{ln}\,\text{cosh}[Wx+b]_i.
    \label{NN_eqn}
\end{equation}
It is important to choose differentiable activation functions for deriving the real-time evolution, which relies on the wave function being differentiable at every point of the variational manifold \cite{schmitt_quantum_2020}. Thus, instead of the reLU($x$) activation function (nondifferentiable at $x=0$), that was used previously to study the ground states of quantum skyrmions \cite{joshi_ground_2023}, in this work, we use an RBM with ln cosh($x$) activation function. The loss function $\mathcal{L}_0(\boldsymbol{\theta})$ for ground state calculations is the energy of the Hamiltonian $H_0$ which is minimized with respect to the variational parameters $\boldsymbol{\theta}$
\begin{equation}
    \mathcal{L}_0(\boldsymbol{\theta})=\left<\psi_{\boldsymbol{\theta}}\right|H_0\left|\psi_{\boldsymbol{\theta}}\right>.
\end{equation}
We use Adam \cite{kingma_adam_2017} as the optimizer and Markov chain Monte Carlo to sample the input spin configurations. 

For the real-time evolution, we use the time-dependent variational principle, which corresponds to the time dependence of the variational parameters, $\boldsymbol{\theta}(t)$. Given a Hamiltonian $H$, at each time step, the loss function is the distance between an infinitesimal time evolved state and the state at time $t^\prime=t+\delta t$
\begin{align}
\nonumber    \mathcal{L}_{t^\prime}(\boldsymbol{\theta}(t^\prime),\boldsymbol{\theta}(t))=\text{dist}\left(\big|\psi_{\boldsymbol{\theta}(t^\prime)}\big>,e^{-iH\delta t}\left|\psi_{\boldsymbol{\theta}(t)}\right>\right),\\
    =\text{dist}\left(\big|\psi_{\boldsymbol{\theta}(t^\prime)}\big>,-i\delta tH\left|\psi_{\boldsymbol{\theta}(t)}\right>\right) + \mathcal{O}(\delta t^2).
    \label{distance}
\end{align}
The parameters $\boldsymbol{\theta}(t)$ are known, and parameters $\boldsymbol{\theta}(t^\prime)$ are to be determined by minimizing the loss function above. We use the Fubini-Study metric as the distance between two wave functions $\left|\psi\right>$ and $\left|\phi\right>$,
\begin{equation}
    \text{dist}(\left|\psi\right>,\left|\phi\right>)=\text{cos}^{-1}\left(\sqrt{\frac{\left<\psi|\phi\right>\left<\phi|\psi\right>}{\left<\psi|\psi\right>\left<\phi|\phi\right>}}\right)^2.
    \label{Fubini-Study}
\end{equation}
Minimizing Eq.~\eqref{distance} results in an equation for the time derivative of the variational parameters $\Dot{\boldsymbol{\theta}}$ \cite{carleo_solving_2017}
\begin{equation}
    \boldsymbol{S}\Dot{\boldsymbol{\theta}}=-i\boldsymbol{F}.
    \label{qgt}
\end{equation}
Here, $S$ is the quantum geometric tensor and $F$ is the force vector defined as (dropping the $t$ dependence of $\boldsymbol{\theta}(t)$ for readability),
\begin{align}
    S_{ij}&=\frac{\left<\partial_{\boldsymbol{\theta}_i}\psi_{\boldsymbol{\theta}}|\partial_{\boldsymbol{\theta}_j}\psi_{\boldsymbol{\theta}}\right>}{\left<\psi_{\boldsymbol{\theta}}|\psi_{\boldsymbol{\theta}}\right>}-\frac{\left<\partial_{\boldsymbol{\theta}_i}\psi_{\boldsymbol{\theta}}|\psi_{\boldsymbol{\theta}}\right>}{\left<\psi_{\boldsymbol{\theta}}|\psi_{\boldsymbol{\theta}}\right>} \frac{\left<\psi_{\boldsymbol{\theta}}|\partial_{\boldsymbol{\theta}_j}\psi_{\boldsymbol{\theta}}\right>}{\left<\psi_{\boldsymbol{\theta}}|\psi_{\boldsymbol{\theta}}\right>},\\
    F_i&=\frac{\left<\partial_{\boldsymbol{\theta}_i}\psi_{\boldsymbol{\theta}}\right|H\left|\psi_{\boldsymbol{\theta}}\right>}{\left<\psi_{\boldsymbol{\theta}}|\psi_{\boldsymbol{\theta}}\right>}-\frac{\left<\partial_{\boldsymbol{\theta}_i}\psi_{\boldsymbol{\theta}}|\psi_{\boldsymbol{\theta}}\right>}{\left<\psi_{\boldsymbol{\theta}}|\psi_{\boldsymbol{\theta}}\right>} \frac{\left<\psi_{\boldsymbol{\theta}}\right|H\left|\psi_{\boldsymbol{\theta}}\right>}{\left<\psi_{\boldsymbol{\theta}}|\psi_{\boldsymbol{\theta}}\right>}.
\end{align}
Both $S$ and $F$ are computed by estimating the expectation values in the Monte Carlo scheme. Finally, Eq.~\eqref{qgt} is integrated with the classic Runge-Kutta method to obtain the parameters $\boldsymbol{\theta}(t^\prime)$. We use NetKet to implement the RBM, VMC, and t-VMC algorithms \cite{carleo_netket_2019,vicentini_netket_2022,mpi4jax:2021}. Details about the hyperparameters used are given in Appendix A.

\section{Ground state}
\label{ground_state}

A quantum skyrmion lattice (QSL) is the ground state of the Hamiltonian in Eq.~\eqref{Ham} for large DMI and finite anisotropy and magnetic field if the lattice size is large enough to accommodate the QSL, consistent with previous findings \cite{haller_quantum_2022,joshi_ground_2023}. The spin expectation values $\left<\boldsymbol{S}\right>=\left<\boldsymbol{\sigma}\right>/2$ in Fig.~\ref{gs_fig}(a) show two quantum skyrmions in the ground state, encircled by dashed lines. As this is a quantum spin model, the lengths of the spins are not normalized due to quantum fluctuations, and thus $|\left<\boldsymbol{S}\right>|<1/2$. The ground state energy $E_0$ minimization plot for the RBM used to describe the QSL in a $9\times9$ lattice is shown in Fig.~\ref{gs_fig}(b), with the energy variance $\left<\psi_{\boldsymbol{\theta}}\right|(H_0-E_0)^2\left|\psi_{\boldsymbol{\theta}}\right>$ in the inset. Here, the Hamiltonian parameters are $D=J$, $A=0.5J$, and $B^z=J$. The RBM converges to a QSL as the variance vanishes.

To characterize the quantum skyrmions, we calculate the local skyrmion density for the nearest neighbor spins $i$, $j$, and $k$ forming an elemental triangle $\Delta$ as \cite{berg_definition_1981,siegl_controlled_2022,peters_quantum_2023}
\begin{align}
    \Omega_{\Delta}=\frac{1}{2\pi}\text{atan2}&({\textbf{n}_i\cdot(\textbf{n}_j\times\textbf{n}_k)},\\
    &{1+\textbf{n}_i\cdot\textbf{n}_j+\textbf{n}_j\cdot\textbf{n}_k+\textbf{n}_k\cdot\textbf{n}_i}).
    \label{sk_density}
\end{align}
Here, we use the normalized spin expectation values $\textbf{n}_i=\left<\boldsymbol{S}_i\right>/|\left<\boldsymbol{S}_i\right>|$. The skyrmion number $N$ is given by the sum over all triangles
\begin{equation}
    N=\sum_{\Delta}\Omega_{\Delta}.
    \label{sk_no}
\end{equation}
Using twice the unnormalized spin expectation values in Eq.~\eqref{sk_density} instead of $\textbf{n}$ results in a non-quantized number $Q$, which depends on the length $|\left<\boldsymbol{S}\right>|$ of the spins and is an indicator of the stability of quantum skyrmions, with $Q\rightarrow N$ if and only if the spin expectation values have maximal amplitude \cite{siegl_controlled_2022} and quantum fluctuations completely vanish. For the ground state solution in Fig.~\ref{gs_fig}, we find $N=2$ corresponding to two quantum skyrmions in the ground state and $Q=1.93$, implying that these skyrmions have spin expectation values with magnitude close to $\hbar/2$. 

We note that the existence of a QSL depends not only on the DMI $D$, anisotropy $A$, and external magnetic field $B^z$ but also on the size of the lattice. For square lattices smaller than $9\times9$ spins, we do not find any ground state hosting a quantum skyrmion in the parameter range $0<D/J<2$, $0<A/J<2$ and $0<Bz<2$. While it is possible to obtain a QS ground state in smaller lattices when embedded in a ferromagnetic medium \cite{siegl_controlled_2022,joshi_ground_2023}, in the presence of periodic boundaries, we only obtain a spin spiral or a ferromagnet as the ground state. For larger lattice sizes up to $13\times13$, we also obtain a QSL as the ground state with $N=2$ for large DMI.

Next, we study entanglement in the QSL ground state as previously done for single quantum skyrmions \cite{haller_quantum_2022,joshi_ground_2023}. Using the expectation value of the ``Swap" operator, we calculate the second order Renyi entropy $S_2(\rho_A)$ as a measure of entanglement in quantum skyrmions \cite{hastings_measuring_2010,hibat-allah_recurrent_2020,joshi_ground_2023},
\begin{equation}
    S_2(\rho_A)=-\frac{1}{2}\text{ln}(\text{Tr}(\rho_A^2)).
\end{equation}
Here, $\rho_A$ is the reduced density matrix obtained by dividing the system into subsystems $A$ and $B$ and tracing out the degrees of freedom in subsystem $B$. In all our Renyi entropy calculations, we take subsystem $A$ to be a single spin and partition $B$ to be the remaining spins to see how the spins are entangled with their environment. The heat map in Fig.~\ref{gs_fig}(c) shows the Renyi entropy in the QSL ground state. The entanglement is largest ($S_2(\rho_A)=0.061$) in the region between two skyrmions and smallest ($S_2(\rho_A)=0.004$) around the center of the skyrmions. The entropy for the central spin is nonzero ($S_2(\rho_A)=0.013$), different from the case of a quantum skyrmion embedded in a ferromagnetic medium where the central spin was disentangled from the rest of the lattice. This might be due to different parameter regimes, boundary conditions, and system sizes \cite{haller_quantum_2022,joshi_ground_2023}. As $0\leq S_2(\rho_A) \leq \text{ln}(2)$, the Renyi entropies are still small in the QSL ground state.

\begin{figure*}
	\centering
	\includegraphics[width=\textwidth]{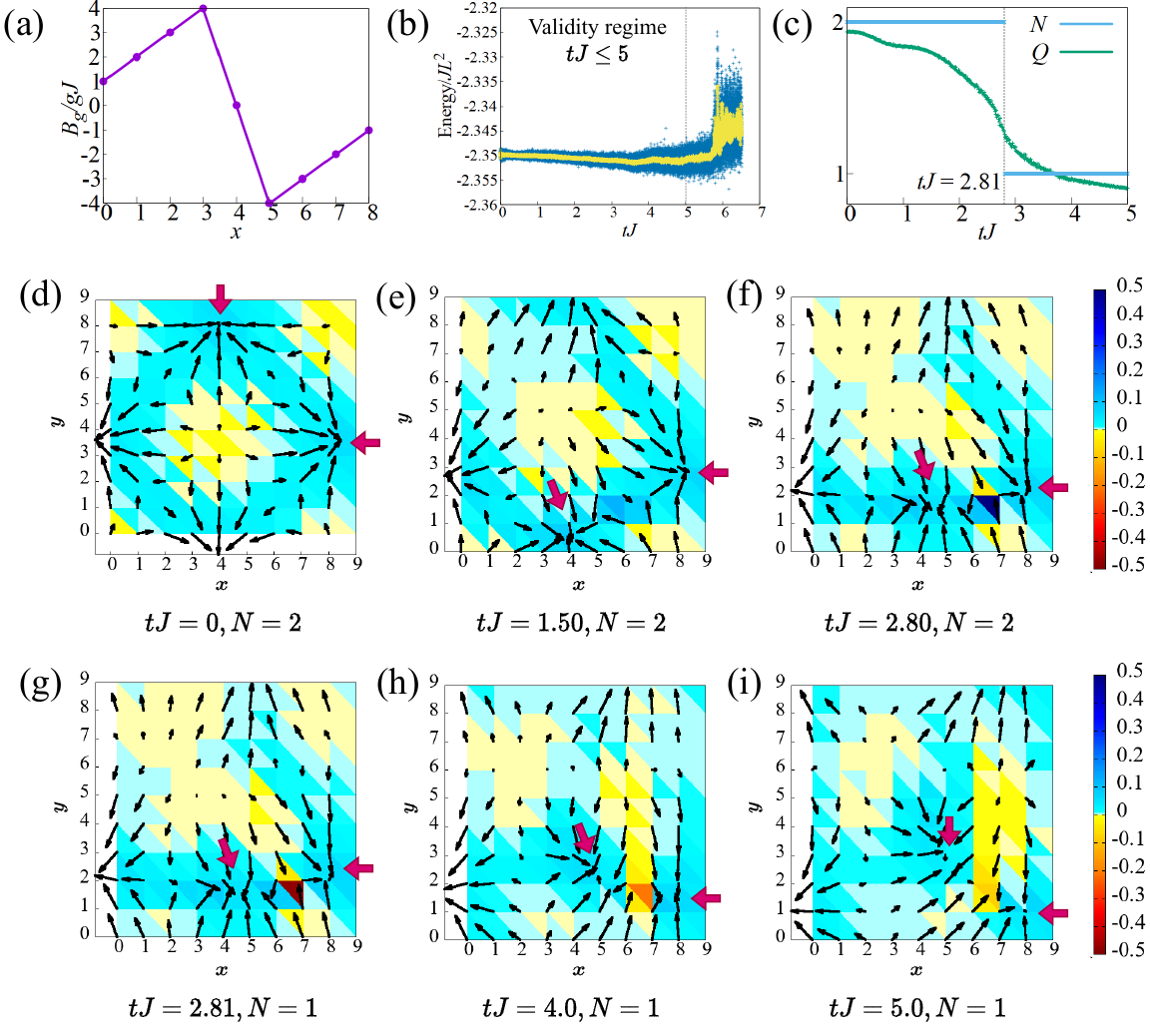}
	\caption{Real-time dynamics of quantum skyrmions. (a) Magnetic field gradient used to move quantum skyrmions (Eq.~\eqref{gradient_field}). The magnetic field points along the $z$-axis and depends on the $x$-coordinate. (b) Time evolution of energy per spin of the quenched Hamiltonian in (Eq.~\eqref{gradient_field}) over time as a quality check for the unitarity of the method. Blue shows the energy at each iteration, and yellow shows the moving average over 30 iterations. (c) Evolution of the normalized skyrmion number $N$ and the unnormalized skyrmion number $Q$ with time. While $Q$ continuously decreases, $N$ is quantized and a transition from $N=2$ to $N=1$ takes place at $tJ=2.81$. (d)-(i) Snapshots of spin expectation values at different times with the skyrmion density $\Omega_\Delta$ in the background. The quantum skyrmions (marked by arrows) move towards each other (d)-(e), interact and an exceptional configuration is formed between $tJ=2.80$ and $tJ=2.81$ (f)-(g), after which one quantum skyrmion decays and an elongated quantum skyrmion remains (h)-(i).}
	\label{te_spins}
\end{figure*}

\section{Dynamics of Quantum Skyrmions}
\label{dynamics_of_quantum_skyrmions}

\begin{figure}
	\centering
	\includegraphics[width=\linewidth]{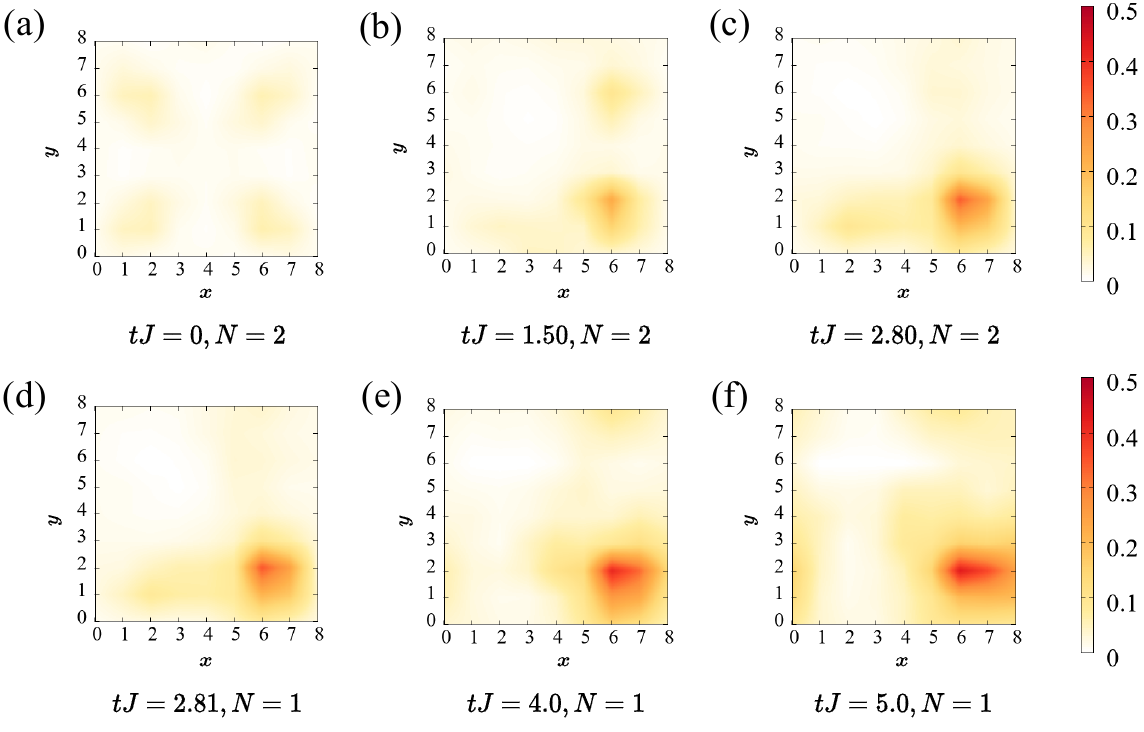}
	\caption{Evolution of Renyi entropy over time. The entropy increases when the quantum skyrmions interact.}
	\label{renyi_entropy}
\end{figure}

In this section, we study the real-time evolution of the QSL ground state after quenching the Hamiltonian with a magnetic field gradient. Magnetic field gradients have been shown to be an effective way of manipulating classical skyrmions and can induce a motion perpendicular to the gradient \cite{zhang_manipulation_2018,komineas_skyrmion_2015,everschor_rotating_2012}. We quench the Hamiltonian in Eq.~\eqref{Ham} with a static, nonuniform magnetic field
\begin{align}
    H_q &= H_0 + \sum_i B^g_i\sigma_i^z, \nonumber\\
    \text{where}, \quad B^g_i &= \begin{cases}
     g(x_i+1) \quad&\text{if}\quad 0\leq x_i<4\\
     0 \quad&\text{if}\quad x_i=4\\
     g(x_i-L) \quad&\text{if}\quad 4<x_i<L\\
    \end{cases}
    \label{gradient_field}
\end{align}
Here, $g$ is the strength of the gradient, $x_i$ is the $x$-coordinate of $i$-th spin, and $x_i=4$ is the $x$-coordinate of the center of one of the skyrmions at $t=0$.
%\textcolor{red}{(The gradient is largest at the center? The gradient seems to be constant (with a jump at the boundary) The magnetic field vanishes at x=xc.)}
The gradient is along the $x$-axis. With this $B^g$, the magnetic field gradient is largest at the center of the middle skyrmion $x_i=4$ and decreases away from it (Fig.~\ref{te_spins}(a)). The speed at which the quantum skyrmions move depends on the gradient, similar to the classical case \cite{komineas_skyrmion_2015}. With this choice of $B^g$, the interaction of quantum skyrmions can be observed in the time scales accessible by our method while maintaining the stability of the nontrivial spin structure. The ground state of the Hamiltonian $H_q$ with magnetic field gradient is a spin spiral phase. Thus, the quench is made from a nontrivial quantum skyrmion phase to a trivial spiral phase. We therefore expect a tendency for the quantum skyrmions to eventually transition to a spiral with $N=0$. With $g=0.2J$, Fig.~\ref{te_spins}(b) shows the evolution of the energy $E_q=\left<\psi_{\boldsymbol{\theta}(t)}\right|H_q\left|\psi_{\boldsymbol{\theta}(t)}\right>$ with time $t$. After the quench, the Hamiltonian is time-independent, the time evolution is unitary, and the energy is supposed to be conserved. While the energy $E_q$ is indeed nearly constant in our simulations, we see that it changes at longer times due to the accumulation of errors \cite{donatella_dynamics_2023,hofmann_role_2022} and we constrain ourselves to the interval $tJ\leq 5$.

The time evolution of spin expectation values is shown in Fig.~\ref{te_spins}(d)-(i) (see Supplementary Information for the video). The color plot in the background shows the local skyrmion density $\Omega_\Delta$ (Eq. \eqref{sk_density}).
The speed of the quantum skyrmions depends on the magnetic field gradient, and as the quantum skyrmion at $x_i=x_c=4$ experiences a larger gradient than the quantum skyrmion at $x_i=8$, it moves faster. The speed of the quantum skyrmions is also proportional to the magnitude of $g$ (not shown here). However, a larger $g$ increases the errors in t-VMC and can even destroy the QSL state. The quantum skyrmions move in a Hall-like motion \cite{zhang_manipulation_2018}, with the velocity perpendicular to the field gradient larger than the velocity parallel to it. The two quantum skyrmions experience opposite magnetic field gradients and move towards each other (Fig.~\ref{te_spins}(e)). The skyrmion density $\Omega_\Delta$ builds up especially for one triangle of spins at $(7,1)$ as the two quantum skyrmions interact. The skyrmion density reaches a maximum of $\Omega_\Delta=0.5$  for this triangle at $tJ=2.80$ Fig.~\ref{te_spins}(f). Then, it passes through an exceptional configuration, where the denominator in Eq.~\eqref{sk_density} changes sign \cite{berg_definition_1981}, and $\Omega_\Delta$ changes from $\Omega_\Delta \approx 0.5$ to $\Omega_\Delta \approx -0.5$ which results in the change of skyrmion number $N$ from $N=2$ to $N=1$ in Fig.~\ref{te_spins}(g). Thus, the quantum skyrmion decay is mediated by exceptional configurations carrying the topological charge of a meron. By this, the two quantum skyrmions merge to an elongated quantum skyrmion (Fig.~\ref{te_spins}(i)). Although $\Omega_\Delta$ changes discontinuously at $tJ=2.81$, the spin expectation values and the wave function do not change discontinuously and the real-time evolution remains valid at this singular point. We also note that we did not observe a dynamical quantum phase transition here, which is accompanied by the nonanalytic behavior of the wave function \cite{vijayan_topological_2023}.

To obtain this decay of quantum skyrmions, it is necessary that the two quantum skyrmions interact. By changing the gradient profile, it becomes possible for the two skyrmions to move in the same direction without interaction. Alternatively, starting with a single skyrmion state (achieved by optimizing the ground state RBM in the presence of large pinning fields such that only one skyrmion remains), the time evolution of a single skyrmion can be obtained. In both cases, we do not observe a quantum skyrmion decay. However, when two quantum skyrmions are driven towards each other, they collide, and this interaction leads to the formation of an exceptional configuration and deletion of a quantum skyrmion.

Finally, let us discuss the evolution of the Renyi entropy with time, shown in Fig.~\ref{renyi_entropy}. At $tJ=0$, the entropy is low and concentrated between the two skyrmions, which is shown in Fig.~\ref{renyi_entropy}(a). As the quantum skyrmions move toward each other, the entropy between them increases, reaching a maximum of $S_2=0.48$ at $tJ=4.50$. The increase in entropy is due to the interaction between the two quantum skyrmions, and it increases continuously, even after one quantum skyrmion decays. The merging of two quantum skyrmions results in large entropy regions, demonstrating the necessity of quantum calculations to capture the correct behavior of this process.

\section{Summary}
\label{summary}

In this work, we studied the ground state properties and real-time evolution of quantum skyrmions. Using variational Monte Carlo with a restricted Boltzmann machine as the variational ansatz, we obtained the ground state of a spin-$1/2$ Heisenberg model in the presence of Dzyaloshinskii-Moriya interaction and Heisenberg anisotropy. The ground state hosts a quantum skyrmion lattice with nonzero Renyi entropy and skyrmion number $N=2$. The Renyi entropy is largest between the two quantum skyrmions. These quantum skyrmions can be manipulated by applying a magnetic field gradient. The quantum skyrmions move in a direction mostly perpendicular to the gradient, with a small parallel component. The velocity of quantum skyrmions depends on the magnitude and direction of the gradient. An exceptional configuration with the topological charge of a meron is formed due to the interaction of the time-evolving quantum skyrmions, resulting in a quantum skyrmion decay as the skyrmion number $N=2$ changes to $N=1$. Thus, neural network quantum states can effectively approximate the real-time evolution of quantum skyrmions and reveal previously unknown quantum phenomena. Stabilizing longer-time dynamics is an interesting aspect for future work.

\begin{acknowledgements}
We thank K. Aoyama for the fruitful discussions. A.J. is supported by the MEXT Scholarship and the Graduate School of Science, Kyoto University, under the Ginfu Fund. A.J. also acknowledges the funding towards this work from the Kyoto University - University of Hamburg fund. R.P. is supported by JSPS KAKENHI No.~JP23K03300. T.P. acknowledges funding by the Deutsche Forschungsgemeinschaft (project no. 420120155) and the European Union (ERC, QUANTWIST, project no. 101039098). Views and opinions expressed are, however, those of the authors only and do not necessarily reflect those of the European Union or the European Research Council. Parts of the numerical simulations in this work have been done using the facilities of the Supercomputer Center at the Institute for Solid State Physics, the University of Tokyo.
\end{acknowledgements}

\appendix
\section{Hyperparameters details}
\label{appendix_A}

For the ground-state calculations, we take the hidden layer density $\alpha=2$. With a larger $\alpha$, energy is slightly improved but the spin expectation values remain unchanged. However, this results in a higher computational cost, especially during real-time evolution. The weights and biases are initialized randomly with a normal distribution having a standard deviation of $0.01$. To optimize the RBM using gradient descent, we use the Adam optimizer with the moments $\beta_1=0.9$ and $\beta_2=0.999$ \cite{kingma_adam_2017}. The learning rate $\eta$ is varied from $\eta=10^{-3}$ to $\eta=10^{-5}$ in the steps of $10^{-1}$ after every $4\times 10^4$ iterations, which can be seen as a sudden drop in the variational energy in Fig.~\ref{gs_fig}(a). Using a stochastic gradient descent optimizer with stochastic reconfiguration \cite{carleo_solving_2017} gives similar results but with increased computational costs. As the Hilbert space is very large, we use Markov chain Monte Carlo to generate samples that are used in the computation of expectation values. The samples are generated by flipping one spin randomly, and the process is repeated $L^2$ times to complete one Monte Carlo sweep. We use $2^{14}$ samples for energy calculation and $2^{17}$ samples for all the other expectation values. 

For the real-time evolution, the optimized RBM is used as the initial state at $tJ=0$. We use a time step of $\delta t=10^{-4}$. At each time step, the quantum geometric tensor $S$ and the forces vector $F$ are computed with $2^{14}$ samples. The equation of motion, Eq.~\eqref{qgt}, can be very unstable due to the presence of noise in the calculation of the matrix $S$ \cite{schmitt_quantum_2020,sinibaldi_unbiasing_2023,hofmann_role_2022}. To improve stability, we add a small shift of $\epsilon=0.01$ to the diagonal elements of the $S$ matrix to regularize the equation of motion. We experimented with different values of $\epsilon$ and found that while the quantum skyrmion motion was similar for all $1.0<\epsilon<10^{-5}$ qualitatively, a smaller $\epsilon$ resulted in unstable energy. Finally, to integrate Eq.~\eqref{qgt}, we use a fourth-order Runge-Kutta integration scheme. Both the ground state optimization and real-time evolution calculations were performed on an NVIDIA A100 GPU.

\begin{figure}[t]
	\centering
	        \includegraphics[width=\linewidth]{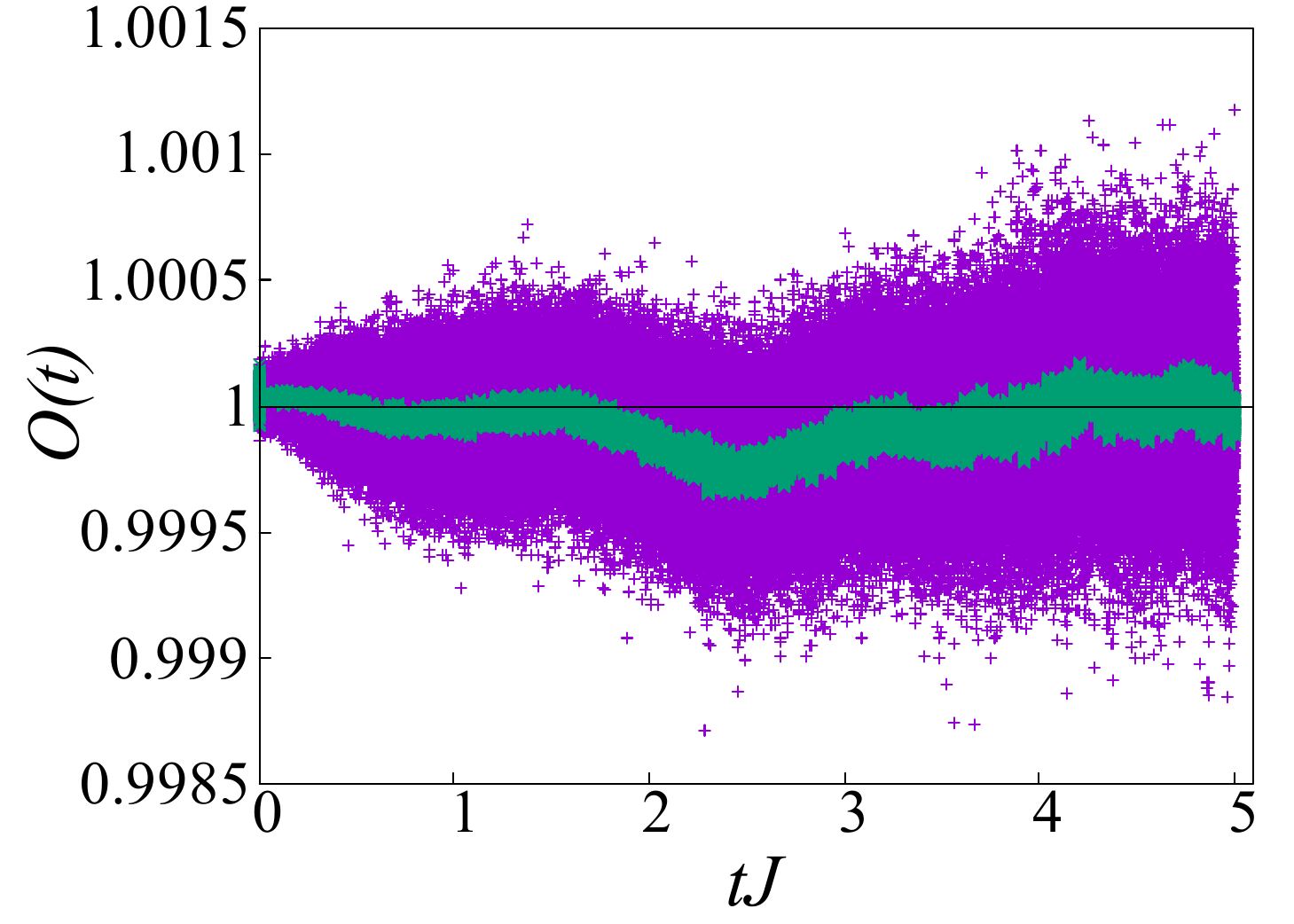}	
	\caption{Overlap between the time evolved state and the state obtained using t-VMC, as defined in Appendix~\ref{appendix_B}. Purple shows the overlap at each time step and green shows the moving average over 100 iterations.}
	\label{residual_fig}
\end{figure}

\section{t-VMC error}
\label{appendix_B}

Here, we discuss the errors in our time evolution calculations. 
%For the results shown in Fig.~\ref{te_spins}, energy evolution in time for $tJ>5$ is shown in Fig.~\ref{residual_fig}(a). The fluctuations in energy start increasing around $tJ=4$ and becomes very large after $tJ>5.6$.
The total error in $\boldsymbol{\theta}$ due to integration of Eq.~\eqref{qgt} using the Runge-Kutta method is of the order $O(\delta t^4)$. The overlap $O(t)= \left<\psi_{\boldsymbol{\theta}(t^\prime)}\right|1-iH\delta t\left|\psi_{\boldsymbol{\theta}(t)}\right>$ is shown in Fig.~\ref{residual_fig}, where $t^\prime=t+\delta t$. Within our approximation, the overlap stays close to unity at each time step.
%The relative residual error $r^2(t)$ gives the error emerging at each time step due to the variational approximation can be calculated using the Fubini-Study metric in Eq.~\eqref{Fubini-Study} as 
%The numerator is minimized to obtain Eq.~\eqref{qgt}, while the denominator is a rescaling factor. The relative residual error for the results shown in Fig.~\ref{te_spins} is shown in Fig~\ref{residual_fig}(b). We see that $r^2(t)$ stays reasonable in our calculations (as compared with Fig. 6 in \cite{carleo_solving_2017}).
%The overlap $\left<\psi_{\boldsymbol{\theta}(t^\prime)}\right|1-iH\delta t\left|\psi_{\boldsymbol{\theta}(t)}\right>$ is also shown in Fig. (not ready yet)

\bibliography{references}

\end{document}